\documentclass{appolb-blank}  %%FRK

\usepackage{graphicx}
\usepackage{latexsym,amssymb,amsmath,amsfonts,epsfig,mathtools}  %%FRK

\newcommand{\beq}{\begin{equation}}   %%FRK
\newcommand{\eeq}{\end{equation}}
\newcommand{\beqa}{\begin{eqnarray}}
\newcommand{\eeqa}{\end{eqnarray}}
\newcommand{\beqNO}{\begin{equation*}}
\newcommand{\eeqNO}{\end{equation*}}
\newcommand{\beqaNO}{\begin{eqnarray*}}
\newcommand{\eeqaNO}{\end{eqnarray*}}
\newcommand{\bsubeqs}{\begin{subequations}}
\newcommand{\esubeqs}{\end{subequations}}
\def\id{\makebox[0.6ex][l]{$1$}{\rm l}}   %%frk from KlinkhamerSchimmel2002

\begin{document}

\eqsec  % uncomment this line to get equations numbered by (sec.num)

\title{\vspace*{-5mm}%
       Towards a numerical solution of the bosonic master-field equation
       of the IIB matrix model
 \\}   %%FRK: extra \\ for nicer lay-out on pp. 1-2

\author{F.R. Klinkhamer
\address{
%\begin{center}
Institute for Theoretical Physics,
Karlsruhe Institute of Technology (KIT),\\
76128 Karlsruhe, Germany\\
\texttt{frans.klinkhamer@kit.edu}
%\end{center}
}\\ %%FRK: extra \\ for nicer lay-out on pp. 1-2
}
\maketitle
\begin{abstract}  %%appolb-blank
A direct algebraic solution has been obtained from the
full bosonic master-field equation of the IIB matrix model
for low dimensionality $D=3$ and small matrix size $N=3$.
A different method is needed for larger values of $D$ and $N$.
Here, we explore an indirect
numerical approach and obtain an approximate
numerical solution for the nontrivial case
$(D,\,N)=(10,\,4)$ with a complex Pfaffian.
We also present a suggestion for numerical calculations at
larger values of $N$.
\end{abstract}

\vspace{-125mm}%%FRK: choose 125mm, to get it at the top of the page
\noindent Acta Phys. Pol. B \textbf{53}, 1--A5  (2022)   
\hfill arXiv:2110.15309
\vspace{+125mm}

\PACS{04.20.Cv, 11.25.-w, 11.25.Yb, 98.80.Bp}  %%frk checked in PACS2010

%%\keywords{General relativity, Strings and branes, M theory,
%%          Origin and formation of the Universe}  %%Epiphany2021

\section{Introduction}
\label{sec:Intro}

It has been argued that
the bosonic large-$N$ master field~\cite{Witten1979,GreensiteHalpern1983}
of the IIB matrix model~\cite{IKKT-1997,Aoki-etal-review-1999}
can give rise to an emergent classical spacetime~\cite{Klinkhamer2020-master}
(see also
Ref.~\cite{Klinkhamer2020-reg-bb-IIB-m-m} for further work on cosmology and
Ref.~\cite{Klinkhamer2021-APPB-review} for a brief review).
The crucial point, now, is that
the bosonic master field is essentially determined by an
algebraic equation and not a differential equation.

In a first paper~\cite{Klinkhamer2021-first-look},
we have considered this algebraic equation with the
effects of fermions removed altogether.
In a subsequent paper~\cite{Klinkhamer2021-sols-D3-N3},
we turned to the full algebraic equation with the
effects of fermions included and were able to obtain
a solution for  dimensionality $D=3$ and matrix size $N=3$.

This result for $(D,\,N)=(3,\,3)$ is certainly gratifying, but
it appears that, with the same type of algebraic method,
progress to larger values of $D$ and $N$ does not seem feasible.
A different method is required to, ultimately,
reach the parameters $D=10$ and $N\gg 1$
of the genuine IIB matrix model.
The present paper explores an indirect numerical approach 
which looks promising
(the meaning of the qualification ``indirect'' will be explained
later).

Before we get started, 
we would like to clarify the scope of the present article.
The numerical results to be given in
Sec.~\ref{sec:Numerical-results} are only indicative.
The importance of the $(D,\,N)=(10,\,4)$ numerical results,
in particular, is to show that the proposed method works.
As such, these numerical results need to be
viewed as a preparation for future work with more powerful computers,
which are needed anyway for reaching larger values of $N$.

%%\newpage%%tmp

\section{Algebraic equation}
\label{sec:Algebraic equation}

As the present paper is solely devoted to obtaining
solutions of a particular algebraic equation,
let us immediately give that equation
(some background on the origin and meaning of the
equation will be given in Sec.~\ref{sec:Background}).
Specifically, this algebraic equation
for $D$ traceless Hermitian matrices $\widehat{a}^{\;\mu}$
of dimension $N \times N$ reads
\begin{subequations}
\label{eq:full-algebraic-equation-defs}
\begin{eqnarray}
\label{eq:full-algebraic-equation}
\hspace*{-6mm}&&
i\,\big(\widehat{p}_{k}-\widehat{p}_{l}\big)\;
\widehat{a}^{\;\mu}_{\;kl}
%&=&
+ \Big[\widehat{a}^{\,\nu},\,\big[\widehat{a}^{\,\nu},\,
\widehat{a}^{\;\mu}\big]\Big]_{kl}
- \frac{1}{\mathcal{P}\left(\widehat{a}\,\right)}\;
\frac{\partial\, \mathcal{P}\left(\widehat{a}\,\right)}
     {\partial\, \widehat{a}^{\;\mu}_{\;lk}}
-\widehat{\eta}^{\;\mu}_{\;kl} = 0\,,
\\[1mm]
\hspace*{-6mm}&&
\mathcal{P}\left(\widehat{a}\,\right) = 
\text{homogeneous polynomial of order $K$}\,,  
\\[1mm]
\label{eq:full-algebraic-equation-def-K}
\hspace*{-6mm}&&
K\equiv \big( D-2\big)\,\big( N^2-1\big) \,,
\\[1mm]
\hspace*{-6mm}&&
D=10\,,\quad N\gg 1\,,
\end{eqnarray}
\end{subequations}
with matrix indices $k,\,l$  running over
$\{  1,\,  \ldots \,,\,N \}$
and directional indices
$\mu,\, \nu$  running over $\{  1,\,  \ldots \,,\,D \}$,
while $\nu$ in \eqref{eq:full-algebraic-equation}
is implicitly summed over.
The $\widehat{p}_{k}$ in \eqref{eq:full-algebraic-equation}
are uniform random numbers
and the $\widehat{\eta}^{\;\mu}_{\;kl}$
Gaussian random numbers (these numbers
can be fixed once and for all, provided $N$ is large enough;
see Sec.~\ref{subsec:Bosonic master-field equation}
for further details).
There is an explicit expression for the Pfaffian
$\mathcal{P}$ to be discussed later.

The algebraic equation
\eqref{eq:full-algebraic-equation}
is quite a challenge for mathematics and computational science.
But why is that equation also of interest to physics?
Well, the answer is that its solution may contain information about
the emergence of spacetime and the birth of the Universe.
As promised above, we will give some background in the next section,
but the main focus of the present paper is really on 
obtaining solutions of the above algebraic equation.

%%\newpage%%tmp

\section{Background}
\label{sec:Background}

The algebraic equation of interest
arises from the IKKT matrix model~\cite{IKKT-1997}.
That model is also known as the IIB matrix
model~\cite{Aoki-etal-review-1999}, because
the matrix model reproduces the structure
of the light-cone type-IIB superstring field theory.

The IIB matrix model
of Kawai and collaborators~\cite{IKKT-1997,Aoki-etal-review-1999}
has a \emph{finite number} of $N \times N$ traceless Hermitian
matrices: ten bosonic matrices $A^{\mu}$ and
eight fermionic Majorana--Weyl matrices $\Psi_{\alpha}$.
The partition function $Z$ of the
IIB matrix model is defined by the following ``path'' integral:
\beqa
\label{eq:IIB-matrix-model-Z}
&&
Z = \int dA\,d\Psi\;
e^{\displaystyle{-S_\text{bos}(A)-S_\text{ferm}(\Psi,\,A)}}.
\eeqa
Here, the bosonic action $S_\text{bos}(A)$ is quartic in $A$
(the trace of the square of Yang--Mills-type commutators)
and the fermionic action $S_\text{ferm}(\Psi,\,A)$ is quadratic
in $\Psi$ and linear in $A$ (a Dirac-type term without derivatives
but with Yang--Mills-type commutators).
In a symbolic notation, the fermionic action reads
\beqa
\label{eq:Sferm}
&&
S_\text{ferm} = \overline{\Psi}\, \mathcal{M}(A)\,\Psi\,.
\eeqa
The precise definition of the measure in \eqref{eq:IIB-matrix-model-Z}
and further details are summarized
in Ref.~\cite{Klinkhamer2021-sols-D3-N3}.

The fermionic matrices $\Psi$ in \eqref{eq:IIB-matrix-model-Z}
can be integrated out exactly (Gaussian integrals) 
and there results a Pfaffian: 
\beqa
\label{eq:Pfaffian-calP}
&&
\text{Pf}\big[\mathcal{M}(A)\big]\equiv \mathcal{P}(A)\,,
\eeqa
with further details collected in App.~\ref{app:Pfaffian-D10-Ngeq2}.
The final expression for the partition function
then reads
\begin{subequations}\label{eq:IIB-matrix-model-Z-Pfaffian-Seff}
\beqa
\label{eq:IIB-matrix-model-Z-Pfaffian}
\hspace*{0mm}
&&
Z =\int dA\;\mathcal{P}(A)\;
e^{\displaystyle{-S_\text{bos}(A)}}
=
\int dA\;e^{\displaystyle{-S_\text{eff}(A)}}\,,
\\[2mm]
\label{eq:Seff}
&&
S_\text{eff}(A)
=
S_{\text{bos}}(A)- \log\,\mathcal{P}(A)\,.
\eeqa
\end{subequations}
For the bosonic observable
\beqa
\label{eq:IIB-matrix-model-w-observable}
&&
w^{\mu_{1} \,\ldots\, \mu_{m}}
\equiv
\text{Tr}\,\big( A^{\mu_{1}} \cdots\, A^{\mu_{m}}\big)\,,
\eeqa
and arbitrary strings thereof, the expectation values are defined by 
the same integral as in \eqref{eq:IIB-matrix-model-Z-Pfaffian}: 
\beqa \label{eq:IIB-matrix-model-w-product-vev}
\hspace*{-4mm}&&
\langle
w^{\mu_{1}\,\ldots\,\mu_{m}}\:w^{\nu_{1}\,\ldots\,\nu_{n}}\, \cdots\,
w^{\omega_{1}\,\ldots\,\omega_{z}}
\rangle
\nonumber\\   \hspace*{-0mm}&&
=
\frac{1}{Z} \int dA\,
\big(w^{\mu_{1}\,\ldots\,\mu_{m}}\:w^{\nu_{1}\,\ldots\,\nu_{n}}\, \cdots\,
w^{\omega_{1}\,\ldots\,\omega_{z}}\big)\,
e^{\displaystyle{-\,S_\text{eff}}}\,.
\eeqa

At this moment, we can make a trivial but important observation:
the matrices of the IIB matrix model \eqref{eq:IIB-matrix-model-Z}
have no spacetime dependence (for this reason, we have used
quotation marks in the terminology ``path'' integral).
In fact, the IIB matrix model just gives numbers,
$Z$ and the expectation values $\langle w\:w\,\cdots\,w\rangle$, while
the matrices $A^{\mu}$ and $\Psi_{\alpha}$
in the ``path'' integral are merely integration variables.
In addition, there is no small dimensionless parameter
to motivate a saddle-point approximation.
So, how does the classical spacetime emerge?
Recently, we have suggested~\cite{Klinkhamer2020-master}
to revisit an old idea,
the large-$N$ master field of Witten~\cite{Witten1979,GreensiteHalpern1983},
for a possible origin of classical spacetime
in the context of the IIB matrix model.

According to Witten~\cite{Witten1979}, the large-$N$ factorization
of the expectation values \eqref{eq:IIB-matrix-model-w-product-vev}
implies that the path integrals are
saturated by a \emph{single} configuration,
the so-called master field $\widehat{A}^{\,\mu}$.
To leading order in $N$, the expectation values are then given by
\bsubeqs \label{eq:IIB-matrix-model-w-product-vev-from-master-field}
\beqa
\hspace*{-10mm}
&&\langle
w^{\mu_{1}\,\ldots\,\mu_{m}}\,w^{\nu_{1}\,\ldots\,\nu_{n}} \cdots\,
w^{\omega_{1}\,\ldots\,\omega_{z}} \rangle
\,\stackrel{N}{=}\,
\widehat{w}^{\,\mu_{1}\,\ldots\,\mu_{m}}\,
\widehat{w}^{\,\nu_{1}\,\ldots\,\nu_{n}} \cdots\,
\widehat{w}^{\,\omega_{1}\,\ldots\,\omega_{z}},
\\[2mm]
\hspace*{-10mm}
&&
\widehat{w}^{\,\mu_{1}\,\ldots\, \mu_{m}}
\equiv
\text{Tr}\,\big( \widehat{A}^{\,\mu_{1}} \cdots\, \widehat{A}^{\,\mu_{m}}\big)\,.
\eeqa
\esubeqs
Hence, we do not have to perform the integral on the right-hand side 
of \eqref{eq:IIB-matrix-model-w-product-vev}:
we only need ten traceless Hermitian matrices
$\widehat{A}^{\,\mu}$ to get \emph{all} these expectation values
from the simple procedure of replacing each $A^{\,\mu}$
in the observables by the corresponding $\widehat{A}^{\,\mu}$.

Now, the meaning of our previous suggestion is clear:
classical spacetime may reside in the bosonic
master-field matrices $\widehat{A}^{\,\mu}$
of the IIB matrix model. The heuristics
is as follows~\cite{Klinkhamer2021-APPB-review}:
\begin{itemize}
  \item
The expectation values
$\left\langle w^{\mu_{1}\,\ldots\,\mu_{m}}
\,\cdots\,w^{\omega_{1}\,\ldots\,\omega_{z}} \right\rangle$
from \eqref{eq:IIB-matrix-model-w-product-vev}
correspond to an infinity of real numbers
and give a large part of the \emph{information content}
of the IIB matrix model
(but, obviously, not all the information).
  \item\vspace*{-0mm}
That \emph{same} information is encoded
in the master-field matrices $\widehat{A}^{\,\mu}$,
which, to leading order in $N$, give precisely the same numbers
by the products $\widehat{w}^{\,\mu_{1}\,\ldots\,\mu_{m}}\, \cdots\,$
$\widehat{w}^{\,\omega_{1}\,\ldots\,\omega_{z}}$,
where $\widehat{w}$ is simply the observable $w$
evaluated for $\widehat{A}$.
  \item\vspace*{-0mm}
From these master-field matrices $\widehat{A}^{\,\mu}$,
it is possible to \emph{extract} the points and the metric of  
an emergent classical spacetime
(as mentioned before, the original matrices $A^{\,\mu}$
are merely integration variables).
\end{itemize}

Assuming that the matrices $\widehat{A}^{\,\mu}$ of the
IIB-matrix-model master field are known
and that they are approximately band-diagonal (as suggested
by the numerical results of
Ref.~\cite{KimNishimuraTsuchiya2012,NishimuraTsuchiya2019,%
Anagnostopoulos-etal-2020} and references therein),
it is relatively easy~\cite{Klinkhamer2020-master}
to extract a discrete set of spacetime points
$\{\widehat{x}^{\,\mu}_{k}\}$
and an interpolating inverse metric $g^{\mu\nu}(x)$.
The metric $g_{\mu\nu}(x)$ is obtained as matrix inverse of $g^{\mu\nu}(x)$.

But, instead of just assuming the matrices $\widehat{A}^{\,\mu}$,
we want to calculate them. And, for that, we need an equation.

%%\newpage%%tmp

\section{Master-field equation: General discussion}
\label{sec:Master-field-equation-general-discussion}

\subsection{Preliminary remarks}
\label{subsec:Preliminary-remarks}

Let us start with some good news:
the master-field equation has already been established,
nearly 40 years ago,
by Greensite and Halpern~\cite{GreensiteHalpern1983}.
They write in the first line of their abstract:
``We derive an exact algebraic (master) equation for the
  euclidean master field of any
  large-$N$ matrix theory, including quantum chromodynamics.''
Now, ``any'' means ``any'' and we may as well consider
the large-$N$ IIB matrix theory~\cite{IKKT-1997,Aoki-etal-review-1999}.

A side remark is that the sentence quoted above is,
perhaps, a little bit too general. For example, an obvious restriction
would be the restriction to any \emph{consistent} large-$N$ matrix theory,
where the additional adjective  ``consistent'' 
implies that the theory makes sense physically (being, for example,  
causal and reflection-positive/unitary).
In any case, the  IIB matrix model is certainly a good candidate
for a consistent large-$N$ matrix theory.

%%%%\newpage%%tmp

\subsection{Bosonic master-field equation}
\label{subsec:Bosonic master-field equation}

Building on the work by Greensite and Halpern~\cite{GreensiteHalpern1983},
we then have the IIB-matrix-model bosonic master field
in a ``quenched'' form~\cite{Klinkhamer2020-master}: 
\vspace*{-0mm}
\bsubeqs
\label{eq:IIB-matrix-model-master-field-algebraic-equation}
\beqa
\label{eq:IIB-matrix-model-master-field}
&&
\widehat{A}^{\;\mu}_{\;kl} =
e^{\displaystyle{i\,\widehat{p}_{k}\,\tau_\text{eq}}}
\;\,\widehat{a}^{\;\mu}_{\;kl}\;\,
e^{\displaystyle{-i\,\widehat{p}_{l}\,\tau_\text{eq}}}\,.
\vspace*{-0mm}
\eeqa
Here, the $\widehat{p}_{k}$ are dimensionless random constants
(see below) and the dimensionless time $\tau_\text{eq}$
must have a sufficiently large value
in order to represent an equilibrium situation
($\tau$ is the fictitious Langevin
time of the stochastic-quantization procedure).
The $\tau$-independent matrix $\widehat{a}^{\;\mu}$
on the right-hand side of \eqref{eq:IIB-matrix-model-master-field}
solves the following algebraic equation~\cite{Klinkhamer2020-master}:
\begin{eqnarray}
\label{eq:IIB-matrix-model-algebraic-equation}
&&
i\,\big(\widehat{p}_{k}-\widehat{p}_{l}\big)\;
\widehat{a}^{\;\mu}_{\;kl}
+ \frac{\partial\, S_\text{eff\,}\left(\widehat{a}\,\right)}
       {\partial\, \widehat{a}_{\mu\;lk}}
-\widehat{\eta}^{\;\mu}_{\;kl} = 0\,,
\vspace*{-0mm}
\end{eqnarray}
\esubeqs
in terms of the effective action $S_\text{eff\,}\left(\widehat{a}\,\right)$
from \eqref{eq:Seff}, the master momenta
$\widehat{p}_{k}$ (real uniform random numbers), and
the master noise  matrices $\widehat{\eta}^{\;\mu}_{\;kl}$
[using real Gaussian random numbers for the
corresponding coefficients $\widehat{\eta}^{\;\mu}_{\;c}$
with Lie-algebra index $c$, where the expansion is similar to
\eqref{eq: master-field-matrix-expansion}
in App.~\ref{app:Pfaffian-D10-Ngeq2}\,].

Further details on the interpretation
of \eqref{eq:IIB-matrix-model-master-field}
and \eqref{eq:IIB-matrix-model-algebraic-equation}
can be found in Sec.~3.2 of Ref.~\cite{Klinkhamer2021-sols-D3-N3}.
At this moment, we like to emphasize one point   
(already briefly mentioned in Sec.~\ref{sec:Algebraic equation}):
the random numbers
$\widehat{p}_{k}$ and $\widehat{\eta}^{\;\mu}_{\;kl}$
can be \emph{fixed once and for all} (assuming that $N$ is large enough)
and these random numbers are called the
\emph{master} momenta and noise.
This observation implies that, at large $N$, it suffices
to solve the algebraic equation
\eqref{eq:IIB-matrix-model-algebraic-equation}
with a \emph{particular} realization of random numbers
$\widehat{p}_{k}$ and $\widehat{\eta}^{\;\mu}_{\;kl}$,
and this is the strategy employed in our previous
papers~\cite{Klinkhamer2021-first-look,Klinkhamer2021-sols-D3-N3}
and the present one.

The algebraic equation \eqref{eq:IIB-matrix-model-algebraic-equation}
is, of course, precisely \eqref{eq:full-algebraic-equation}.

%%\newpage%%tmp

\subsection{Simplified equation}
\label{subsec:Simplified-equation}

The algebraic equation \eqref{eq:full-algebraic-equation}
is truly formidable and it makes sense
to first consider the simplified equation 
without Pfaffian term~\cite{Klinkhamer2021-first-look}: 
\beqa
\label{eq:simplified-algebraic-equation}
&&
i\,\big(\widehat{p}_{k}-\widehat{p}_{l}\big)\;
\widehat{a}^{\;\mu}_{\;kl}
+
\Big[\widehat{a}^{\,\nu},\,\big[\widehat{a}^{\,\nu},\,
\widehat{a}^{\;\mu}\big]\Big]_{kl}
-\widehat{\eta}^{\;\mu}_{\;kl}=0\,.
\eeqa
The matrices $\widehat{a}^{\;\mu}$ are
$N \times N$ traceless Hermitian matrices
and the number of variables is
\beqa
\label{eq:Nvar}
&&
N_\text{var} = D\,\big(N^2-1\big)\,,
\eeqa
which grows rapidly with increasing $N$.
The simplified equation \eqref{eq:simplified-algebraic-equation}
is essentially a cubic polynomial.
Remark also that
the simplified  equation \eqref{eq:simplified-algebraic-equation}
differs from the one in Ref.~\cite{Klinkhamer2021-first-look}
by the sign of the double commutator, but this can be compensated
by a redefinition of $\widehat{p}_{k}$ and $\widehat{\eta}^{\;\mu}_{\;kl}$.

It appears impossible to obtain a \emph{general} analytic solution 
of \eqref{eq:simplified-algebraic-equation} in
terms of the master constants
$\widehat{p}_{k}$ and  $\widehat{\eta}^{\;\mu}_{\;kl}$.
Instead, we have obtained
solutions~\cite{Klinkhamer2021-first-look}
for $(D,\,N)=(2,\,4)$ and $(D,\,N)=(2,\,6)$
by taking a \emph{particular} realization of the random master constants 
(other realizations give similar results).
As our focus will be on $N=4$ in the following, we briefly review  
here the results of the
simplified equation \eqref{eq:simplified-algebraic-equation}
for $(D,\,N)=(2,\,4)$.

\begin{figure}[t]
%\vspace*{-0mm}
\begin{center}
\hspace*{0mm} %%renamed --> num-sol-master-FIG1-v3.eps --> num-sol-master-FIG1-v4.eps
\includegraphics[width=0.7\textwidth]{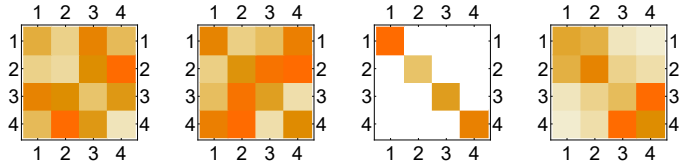}
%{num-sol-bosonic-master-field-eq-FIG1-v1.eps}
%{plotABSahatmuahatmuprimeD2N4simpl-23oct2021.eps}
\end{center}\vspace*{-0mm}
\caption{(color) From left to right, density plots of the matrices
$\text{Abs}\left[\widehat{a}^{\,1}_{\alpha\text{-sol}}\right]$,
$\text{Abs}\left[\widehat{a}^{\,2}_{\alpha\text{-sol}}\right]$,
$\text{Abs}\left[\widehat{a}^{\;\prime\,1}_{\alpha\text{-sol}}\right]$,
and $\text{Abs}\left[\widehat{a}^{\;\prime\,2}_{\alpha\text{-sol}}\right]$,
as obtained in Sec.~4.1 of Ref.~\cite{Klinkhamer2021-first-look}
from the simplified algebraic equation \eqref{eq:simplified-algebraic-equation}
for $(D,\,N)=(2,\,4)$ and
a particular choice (labelled ``$\alpha$'')
of master momenta and master noise.
}
\label{fig:plotABSahatmuahatmuprimeD2N4simpl}
\vspace*{0mm}
\end{figure}

Taking a particular realization (the ``alpha-realization'')
of the 4 pseudorandom numbers for the master momenta
and the 30 pseudorandom numbers for the master noise
(they are given by Eqs.~(16abc) in Ref.~\cite{Klinkhamer2021-first-look},
with overall minus signs added),
we have obtained an explicit solution
$\widehat{a}^{\,1}_{\alpha\text{-sol}}$
and $\widehat{a}^{\,2}_{\alpha\text{-sol}}\,$
as given by Eqs.~(17ab) in Ref.~\cite{Klinkhamer2021-first-look}.
For the absolute values of these matrix entries,
the density plots are shown in the first two panels
of Fig.~\ref{fig:plotABSahatmuahatmuprimeD2N4simpl}.
There is no obvious band-diagonal structure.

Next, change the basis, in order to
diagonalize and order the $\mu=1$ matrix.
This produces the matrices
$\widehat{a}^{\;\prime\,1}_{\alpha\text{-sol}}$
and $\widehat{a}^{\;\prime\,2}_{\alpha\text{-sol}}$
as given by Eqs.~(21ab) in Ref.~\cite{Klinkhamer2021-first-look}.
For the absolute values of these new matrix entries,
the density plots are shown in the last two panels
of Fig.~\ref{fig:plotABSahatmuahatmuprimeD2N4simpl}.
There appears a clear band-diagonal structure   
in $\widehat{a}^{\;\prime\,2}_{\alpha\text{-sol}}$,   
which is a nontrivial result.

As mentioned in Sec.~\ref{sec:Background},
the diagonal/band-diagonal structure
of the master-field matrices allows for the extraction
of a classical spacetime, but it remains to demonstrate
that dynamical fermions do not spoil this structure.

%%\newpage%%tmp

\subsection{Full equation rewritten with a trace term}
\label{subsec:Full-equation-rewritten}

We now look for solutions of
the full bosonic master-field equation
\eqref{eq:full-algebraic-equation},
with dynamic fermions included,
but, first, only for rather small values of $D$ and $N$.

The Pfaffian is a $K$-th order homogenous polynomial,
denoted symbolically by $P_{K}(A)$, with $K = (D-2)\,(N^2-1)$.
Further details are given in App.~\ref{app:Pfaffian-D10-Ngeq2}.
The basic structure of the algebraic
equation \eqref{eq:full-algebraic-equation} is then as follows:
\vspace*{-0mm}
\beqa
\label{eq:algebraic-equation-with-D-structure}
&&
P_{1}^{(\,\widehat{p}\,)}\left(\widehat{a}\,\right)
+
P_{3}\left(\widehat{a}\,\right)
+\frac{P_{K-1}\left(\widehat{a}\,\right)}{P_{K}\left(\widehat{a}\,\right)}
+P_{0}^{(\,\widehat{\eta}\,)}\left(\widehat{a}\,\right)
=0\,,
\vspace*{-0mm}
\eeqa
where the suffixes on $P_{0}$ and $P_{1}$
indicate their respective dependence on
the master noise $\widehat{\eta}^{\;\mu}_{kl}$ and
the master momenta $\widehat{p}_{k}$.
Multiplying \eqref{eq:algebraic-equation-with-D-structure} by
$P_{K}\left(\widehat{a}\,\right)$, we get a polynomial equation
of order $K+3$ with the following structure:%
\beqa
\label{eq:polynomial-equation}
&&
P_{K+1}^{(\,\widehat{p}\,)}\left(\widehat{a}\,\right)
+
P_{K+3}\left(\widehat{a}\,\right)
+P_{K-1}\left(\widehat{a}\,\right)
+P_{K}^{(\,\widehat{\eta}\,)}\left(\widehat{a}\,\right)
=0\,.
\vspace*{-0mm}
\eeqa

For the case $\{ D,\,  N \}  = \{ 3,\,  3 \}$,
there is an explicit algebraic result
for the Pfaffian~\cite{KrauthNicolaiStaudacher1998}.
Taking a particular realization of the
random constants (other realizations give similar results),
we have established~\cite{Klinkhamer2021-sols-D3-N3}
the existence of several solutions of
the full bosonic master-field equation
\eqref{eq:full-algebraic-equation}.
Moreover, there is a mild diagonal/band-diagonal structure,
but the value $N=3$ is too small for definitive statements.

These $(D,\,N)=(3,\,3)$ results
were obtained by a direct algebraic calculation
and it would seem difficult to go to larger values of $D$ and $N$.
But perhaps there can be progress with a numerical approach.
The idea~\cite{Nishimura-private}
is to use the fact that the square of the Pfaffian of the  skew-symmetric matrix
$\mathcal{M}=\mathcal{M}\left(\widehat{a}\,\right)$ equals its determinant,
\label{eq:Pfsquare-equals-det}
\beqa
\big[\text{Pf}(\mathcal{M})\big]^2 = \det \mathcal{M}\,,
\vspace*{-0mm}
\eeqa
so that we can write the variational term in the
algebraic equation \eqref{eq:full-algebraic-equation}
as a trace,
\beqa
\label{eq:varPfoverPf-equals-trace}
\frac{\delta\, \text{Pf}(\mathcal{M})}{\text{Pf}(\mathcal{M})}
=
\frac{1}{2}\;\frac{\delta \det \mathcal{M}}{\det \mathcal{M}}
=
\frac{1}{2}\;\text{Tr}\big[ \mathcal{M}^{-1} \delta\, \mathcal{M}\big]\,,
\vspace*{-0mm}
\eeqa
and this trace can be readily evaluated numerically
(as was done in Ref.~\cite{Anagnostopoulos-etal-2020}).
Just to be clear, it is the third term on the
left-hand side of \eqref{eq:full-algebraic-equation}
that is replaced by a term involving the trace,
according to \eqref{eq:varPfoverPf-equals-trace}.

We will call the resulting calculation an indirect numerical
calculation, where the qualification ``indirect'' is meant to
show that the Pfaffian is not directly considered but rather its relative
variation written as the trace term \eqref{eq:varPfoverPf-equals-trace}
with the matrix $\mathcal{M}_{A,\,B}$ 
from App.~\ref{app:Pfaffian-D10-Ngeq2}.  
First results from this indirect numerical calculation will be
presented in Sec.~\ref{sec:Numerical-results}.

%%\newpage%%tmp

\section{Numerical results from the full algebraic equation}
\label{sec:Numerical-results}

\subsection{Cases $(D,\,N)=(3,\,3)$ and $(D,\,N)=(10,\,3)$}
\label{subsec:Cases-D3-N3-and-D10-N3}

We have performed numerical calculations
of the full algebraic equation \eqref{eq:full-algebraic-equation}
with the trace term \eqref{eq:varPfoverPf-equals-trace} for three cases:
$(D,\,N)=(3,\,3)$, $(D,\,N)=(10,\,3)$, and $(D,\,N)=(10,\,4)$.
Some details of the calculations
are listed in Table~\ref{tab-num-calculation}.

The numerical results of the first two rows in
Table~\ref{tab-num-calculation} were obtained
from the \texttt{NMinimize} routine of \textsc{Mathematica} 12.1
(cf. Ref.~\cite{Wolfram1991}) with the downhill-simplex method
of Nelder and Mead~\cite{NelderMead1965,Press-etal-1992}.
Our main interest is, however, in the $(D,\,N)=(10,\,4)$ case
and we will discuss that case in the next subsection.

\begin{table}[t]
\vspace*{-0mm}
\begin{center}
\caption{Numerical calculations of the full  bosonic master-field
equation \eqref{eq:full-algebraic-equation} using
the identity \eqref{eq:varPfoverPf-equals-trace}.
The number of variables is given by $N_\text{var}=D\,(N^2-1)$
and the order of the Pfaffian by $K = (D-2)\,(N^2-1)$.
\vspace*{1\baselineskip}
}
\label{tab-num-calculation}
\renewcommand{\tabcolsep}{1.0pc}    %% enlarge column spacing
\renewcommand{\arraystretch}{1.5}   %% enlarge line spacing
\begin{tabular}{l|c|c|l}
\hline\hline
  &  $N_\text{var}$
  &  $K$
  &  $\text{calculational details}$\,$^{a}$  %%FRK: footnote does not work
  %
%\footnote{\,for
%a Lenovo T15p notebook with an Intel Core i7-10750H processor
%running \textsc{Mathematica} 12.1.}
%
  \\
\hline\hline
$(D,\,N)=(3,\,3) $   &  24   & 8  &
\text{single-processor},
$\text{O}(\text{hr})$\,$^{b}$  %%FRK: footnote does not work
%
%\footnote{\,reproducing the previous algebraic
%results~\cite{Klinkhamer2021-sols-D3-N3}.}
%
\\
$(D,\,N)=(10,\,3)$   &  80   & 64  &
\text{single-processor}, $\text{O}(\text{day})$\\
$(D,\,N)=(10,\,4)$   &
150\,$^{c}$  %%FRK: footnote does not work
%
%\footnote{\,complex variables in the matrix solution,
%as the Pfaffian $\mathcal{P}\left(\widehat{a}\,\right)$ is complex
%(cf. App.~\ref{app:Pfaffian-D10-Ngeq2}).}
%
& 120 &
\text{6-kernel}, $\text{O}(\text{month})$
\\
\hline\hline
\end{tabular}
\end{center} %%FRK: now BRUTE-FORCE footnotes
$^{a}$\,{\footnotesize for
a Lenovo T15p notebook with an Intel Core i7-10750H processor
(6 kernels, 12}\newline
\phantom{$^{a}$}
{\footnotesize threads, 12 MB cache,
2.6 -- 5.0 GHz clock frequency) 
running \textsc{Mathematica} 12.1.}\newline
$^{b}$\,{\footnotesize reproducing the previous algebraic
results~\cite{Klinkhamer2021-sols-D3-N3}.}\newline
$^{c}$\,{\footnotesize complex variables in the matrix solution, as the Pfaffian $\mathcal{P}\left(\widehat{a}\,\right)$ is complex
(cf. App.~\ref{app:Pfaffian-D10-Ngeq2}).}
\vspace*{0mm}
\end{table}
\vspace*{0mm}

%%\newpage%%tmp

\subsection{Case $(D,\,N)=(10,\,4)$}
\label{subsec:Case-D10-N4}

The downhill-simplex method
used for $(D,\,N)=(3,\,3)$ and $(D,\,N)=(10,\,3)$
is, without further modifications, no longer suitable
for the 300 real variables of the present case,
\beqa
\label{eq:case-D10-N4}
&&
(D,\,N)=(10,\,4)\,.
\eeqa
Instead, we have used a simple random-step
procedure, which could be partially parallelized.
In this way, we have obtained an approximate numerical solution.
A numerical solution is, of course, always ``approximate,''
but occasionally we prefer to emphasize this fact by use
of the adjective.

Before we describe the obtained results, we need
to specify the particular realization (the ``$\kappa$-realization'')
of the pseudorandom numbers entering
the algebraic equation \eqref{eq:full-algebraic-equation}.
As this involves 154 real numbers, the
details are rather cumbersome and are relegated
to App.~\ref{app:Pseudorandom numbers}.
Other realizations of the pseudorandom numbers
(the  $\lambda$, $\mu$, $\nu$, $\ldots$ realizations)
are expected to give similar results, according to our previous
results~\cite{Klinkhamer2021-first-look,Klinkhamer2021-sols-D3-N3}.

The calculation follows the same procedure as in our earlier
work,
namely the numerical minimization of a nonnegative function
$f_\text{penalty}$ that will be defined shortly.
But the calculation for the case \eqref{eq:case-D10-N4}
is hard with many variables and a high-order Pfaffian.
The main difficulty is that,
for a computer as described in Footnote~a of
Table~\ref{tab-num-calculation}, the evaluation of $f_\text{penalty}$
at a single point in the 300-dimensional configuration space
takes a long time, about 90 seconds [which is approximately
ten times more than for the $(D,\,N)=(10,\,3)$ calculation].
Furthermore, the $f_\text{penalty}$ valley appears to be
long, narrow, and winding
(at least, for the used coordinates $\widehat{r}_{\mu}^{\;c}$
and $\widehat{s}_{\mu}^{\;c}$, to be defined shortly).

As the Pfaffian term in the algebraic equation is complex,
we must allow 
for complex variables in the solution: 
\beqa
\label{eq:complex-coefficients}
&&
\widehat{a}_{\mu}^{\;c} =
\widehat{r}_{\mu}^{\;c}+ i\; \widehat{s}_{\mu}^{\;c} \in \mathbb{C}\,, 
\eeqa
with real numbers $\widehat{r}_{\mu}^{\;c}$ and $\widehat{s}_{\mu}^{\;c}$.
In this way, we have obtained a series of
approximate numerical solutions of the
algebraic equation \eqref{eq:full-algebraic-equation}
with the identity \eqref{eq:varPfoverPf-equals-trace}
and the $\kappa$-realization of pseudorandom constants.
A selection of these approximate numerical solutions
is shown in Table~\ref{tab-num-sols},
where the caption defines the function $f_\text{penalty}$,
as well as another diagnostic quantity.
Specifically, we will discuss 
the approximate numerical solution from the last row  of  
Table~\ref{tab-num-sols} and
denote this particular solution by ``$\kappa\text{-num-\underline{sol}}$''.
We, then, have 300 real numbers defining the following matrices:
\beqa
&&
\widehat{a}^{\,\mu}_{\kappa\text{-num-\underline{sol}}}\,,
\;\;\;\text{for}\;\;\; \mu=1,\,\ldots\,,10\,.
\eeqa
With complex coefficients $\widehat{a}_{\mu}^{\;c}$, 
these (approximate) master-field matrices
are no longer Hermitian. The situation is perhaps analogous
to that of complex saddle-points appearing for a real problem.
Our interpretation is that these (approximate) master-field matrices
carry \emph{information} both in their Hermitian and anti-Hermitian
parts.

\begin{table}[t]   
\vspace*{-0mm}
\begin{center}
\caption{Approximate numerical solutions of the full
$(D,\,N)=(10,\,4)$ bosonic master-field
equation \eqref{eq:full-algebraic-equation}
with the identity \eqref{eq:varPfoverPf-equals-trace}
and the pseudorandom constants given in
App.~\ref{app:Pseudorandom numbers}.
The complex residues of the 150 component equations
$\text{eq-}\widehat{a}_{\mu}^{\;c}$
are computed (they all vanish for a perfect solution).
The quantity $\text{MaxAbsRes}$ is the maximum of the absolute values
of these residues
and the function $f_\text{penalty}$ is the sum of their squared
absolute values.
The expression $\text{eq-}\widehat{a}_{\mu}^{\;c}$ follows
from \eqref{eq:full-algebraic-equation}
by performing a matrix multiplication with $t_{c}$,
taking the trace, and multiplying the result by two
[here, $t_{c}$ is the $SU(4)$ generator defined in  
App.~\ref{app:Pfaffian-D10-Ngeq2}
with trace condition \eqref{eq:TraceNormalization}].
}
\vspace*{1\baselineskip}
\label{tab-num-sols}
\renewcommand{\tabcolsep}{1.5pc}    %% enlarge column spacing
\renewcommand{\arraystretch}{1.5}   %% enlarge line spacing
\begin{tabular}{c|c}
\hline\hline
$f_\text{penalty}\equiv \sum \big|\text{eq-}\widehat{a}_{\mu}^{\;c}\big|^{2}$  &
$\text{MaxAbsRes} \equiv
\text{max} \big\{\big|\text{eq-}\widehat{a}_{\mu}^{\;c}\big| \big\}$  \\
\hline\hline
$1375.200$   &  $6.81$  \\
$422.468$    &  $4.56$  \\
$310.932$    &  $3.46$  \\
$209.330$    &  $2.83$  \\
$108.094$    &  $1.83$  \\
%$54.004$    &   $1.48$  \\
%$48.006$       & $1.39$  \\
$39.882$       & $1.15$  \\
\hline\hline
\end{tabular}
\end{center}
\vspace*{-0mm}
\end{table}

We suspect that the Hermitian parts of the
master-field matrices (with real eigenvalues)
contain information about the emerging spacetime~\cite{Klinkhamer2020-master}.
What the information in the anti-Hermitian parts corresponds
to is not clear for the moment (see Sec.~\ref{sec:Discussion}
for a suggestion).

\begin{figure}[t] %%[p]
%\vspace*{-0mm}
\begin{center}
\hspace*{0mm}
\includegraphics[width=0.95\textwidth]{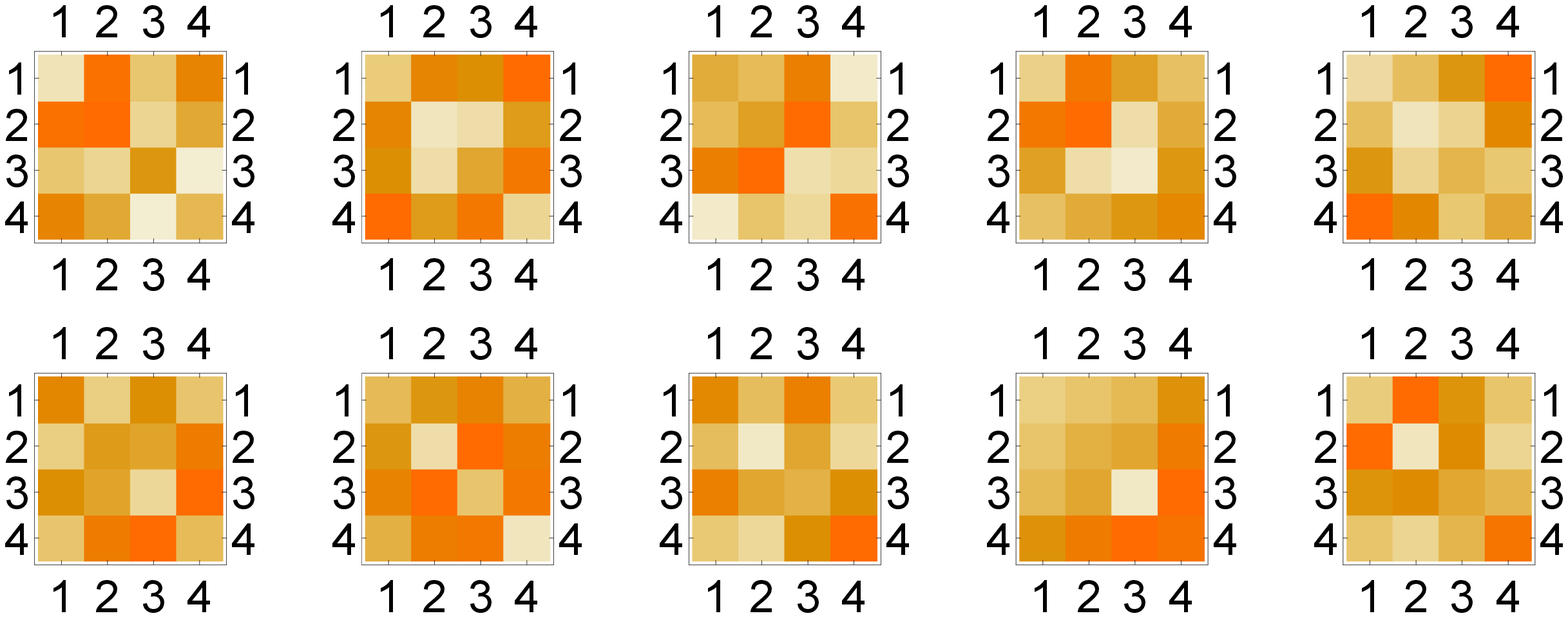}
%%{plotABSahatmu-D10-N4-f39pt882-03jan2022.eps}
%%{num-sol-master-FIG2-v3.eps}
%%{plotABSahatmu-D10-N4-f48pt0058-02dec2021.eps}
%%{plotABSahatmu-D10-N4-f54pt0042-19nov2021.eps}
%%{num-sol-bosonic-master-field-eq-FIG2-v1.eps}
%%{plotABSahatmu-D10-N4-f108pt094-07oct2021.eps}
\end{center}\vspace*{-0mm}
\caption{(color) Density plots of the original
$(D,\,N)=(10,\,4)$ approximate numerical solution
$\text{Abs}\big[\widehat{a}^{\,\mu}_{\kappa\text{-num-\underline{sol}-HERM}}\big]$
with $f_\text{penalty}=39.882$. 
Shown are $\mu=1,\,\ldots\,,5$ on the top row and
$\mu=6,\,\ldots\,,10$ on the bottom row.}
\label{fig:Absahat-f39pt882}
\vspace*{0mm}
\end{figure}

Consider, therefore, the Hermitian parts
\beqa
&&
\widehat{a}^{\,\mu}_{\kappa\text{-num-\underline{sol}-HERM}}
\equiv
\frac{1}{2}\;
\left[
\widehat{a}^{\,\mu}_{\kappa\text{-num-\underline{sol}}} +
\left(\widehat{a}^{\,\mu}_{\kappa\text{-num-\underline{sol}}}\right)^{\dagger}\right]\,.
\eeqa
Calculating the absolute values of these matrix entries,
we observe no obvious band-diagonal structure
(see Fig.~\ref{fig:Absahat-f39pt882}).
\begin{figure}
\vspace*{0mm}
\begin{center}
\hspace*{0mm}
\includegraphics[width=0.95\textwidth]{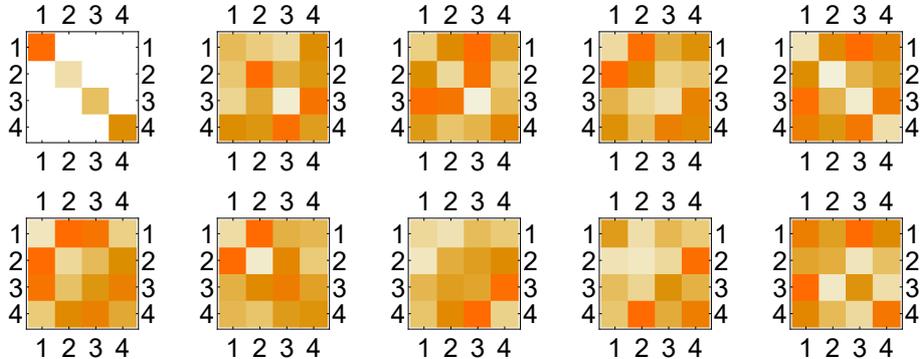}
%%{plotABSahatmuprime-D10-N4-f39pt882-03jan2022.eps}
%%{num-sol-master-FIG3-v3.eps}
%%{plotABSahatmuprime-D10-N4-f48pt0058-02dec2021.eps}
%%{plotABSahatmuprime-D10-N4-f54pt0042-19nov2021.eps}
%%{num-sol-bosonic-master-field-eq-FIG3-v1.eps}
%%{plotABSahatmuprime-D10-N4-f108pt094-07oct2021.eps}
\end{center}\vspace*{-0mm}
\caption{(color) Density plots of the transformed
$(D,\,N)=(10,\,4)$ approximate numerical solution
$\text{Abs}\big[\widehat{a}^{\;\prime\,\mu}_{\kappa\text{-num-\underline{sol}-HERM}}\big]$ 
with $f_\text{penalty}=39.882$.  
Shown are $\mu=1,\,\ldots\,,5$ on the top row and
$\mu=6,\,\ldots\,,10$ on the bottom row.}
\label{fig:Absahatprime-f39pt882}
\vspace*{0mm}
\end{figure}
Now, change the basis, in order to
diagonalize and order the $\mu=1$ matrix.
This gives the matrices
\beqa\label{eq:ahatprime-kappa-num-sol-HERM}
&&
\widehat{a}^{\;\prime\,\mu}_{\kappa\text{-num-\underline{sol}-HERM}}\,,
\;\;\;\text{for}\;\;\; \mu=1,\,\ldots\,,10\,.
\eeqa
Calculating the absolute values of the entries of these transformed matrices,
there is not yet a clear signal for a diagonal/band-diagonal structure:
see Fig.~\ref{fig:Absahatprime-f39pt882} and compare
with Fig.~\ref{fig:plotABSahatmuahatmuprimeD2N4simpl}
without dynamic fermions.
Let us consider the result from
Fig.~\ref{fig:Absahatprime-f39pt882} in somewhat more detail.

The first observation is that
we can perhaps observe a \emph{trend} if we look at the 
improving approximations as listed in the last
five rows of Table~\ref{tab-num-sols}. The
five corresponding density plots of
$\text{Abs}\big[\widehat{a}^{\;\prime\,\mu}_{\kappa\text{-num-\underline{sol}-HERM}}\big]$ 
are shown as Figs. 6--10  in Ref.~\cite{Klinkhamer2021-Corfu-v4}.
In these figures, we can, for example, focus on the
pattern of the $\mu=2$ matrix as it evolves with
improving approximations (decreasing penalty function)
and see that a very weak diagonal/band-diagonal structure
appears at $f_\text{penalty} \sim 300$ and
that this structure more or less stabilizes
at lower values of $f_\text{penalty}$
(note that $N=4$ is still far away from infinity).  
The second observation is that, at best, we can expect  
to find only a \emph{mild} diagonal/band-diagonal structure 
due to the dynamic-fermion effects,
as was previously observed in the algebraic $(D,\,N)=(3,\,3)$
results~\cite{Klinkhamer2021-sols-D3-N3}.

Obviously, we would like to obtain an improved
$(D,\,N)=(10,\,4)$ numerical solution
with $f_\text{penalty} \lesssim 1$, but it is possible that
the basic result as given in Fig.~\ref{fig:Absahatprime-f39pt882}
does not change much.
In any case, the $(D,\,N)=(10,\,4)$ numerical results shown here are,
as mentioned in the last paragraph of Sec.~\ref{sec:Intro},
primarily intended to verify that
the proposed method can be implemented.
The actual $(D,\,N)=(10,\,4)$ results are
only indicative and future work on supercomputers
should go to larger values of $N$
(see Sec.~\ref{sec:Discussion} for further comments).

%%\newpage%%tmp

\section{Discussion}
\label{sec:Discussion}

A direct algebraic solution of the full bosonic master-field
equation  \eqref{eq:full-algebraic-equation}
of the IIB matrix model cannot go much further
than dimensionality $D = 3$ and matrix size $N = 3$.
In the present paper, we have explored an indirect numerical
approach. The qualification ``indirect'' indicates that the
Pfaffian is not directly considered but rather its relative
variation written as the trace term \eqref{eq:varPfoverPf-equals-trace}.
With considerable effort, we have obtained an approximate
numerical solution for the case $(D,\, N) = (10,\, 4)$, which
has a complex Pfaffian.
Having a complex Pfaffian requires a conceptual reinterpretation
of the bosonic master-field,
%with an Hermitian part possibly relevant to
with a Hermitian part possibly relevant to
the emergence of spacetime and
an anti-Hermitian part whose interpretation is
unclear for the moment (perhaps this anti-Hermitian part
is directly or indirectly relevant to the emergence of matter).
Obviously, this is an important open question.

It is, however, difficult to go to $N$ values very much larger than 4,
while keeping $D=10$.
We now have a perhaps somewhat surprising suggestion, namely,
to temporarily leave behind the purely algebraic equation
and to return to the
original Langevin differential equation, but with a special
form of the noise matrices. Specifically, we suggest to use
the quenched-master form for the Langevin noise matrices:
\beqa
\label{eq:quenched-noise}
&&
\eta^{\;\mu}_{\;kl}(\tau)
=
e^{\displaystyle{i\,\widehat{p}_{k}\,\tau}}
\;\,\widehat{\eta}^{\;\mu}_{\;kl}\;\,
e^{\displaystyle{-i\,\widehat{p}_{l}\,\tau}}\,,
\eeqa
with Langevin time $\tau$ and
fixed random numbers $\widehat{p}_{k}$
and $\widehat{\eta}^{\;\mu}_{\;kl}$,
where the matrix indices $k$ and $l$
take values from $\{  1,\,  2,\, \ldots \,,\,N \}$
and the directional index $\mu$
from $\{  1,\,  2,\, \ldots \,,\,D \}$.
See Sec.~3.2 in Ref.~\cite{Klinkhamer2021-sols-D3-N3}
for a brief discussion of the characteristics of these
random numbers $\widehat{p}_{k}$
and $\widehat{\eta}^{\;\mu}_{\;kl}$,
the first having a uniform distribution and the second
$D$ independent Gaussian distributions.
See, furthermore, Sec.~2 of Ref.~\cite{GreensiteHalpern1983}
for the proof that the noise \eqref{eq:quenched-noise}
suffices in the large-$N$ limit (with planar diagrams
dominating).

It appears that the $D = 10$ setup and technology used
in Ref.~\cite{Anagnostopoulos-etal-2020}
[starting from Eqs. (3.2) and (3.3) in that reference],
can be employed, but now with the special noise
matrices \eqref{eq:quenched-noise}.
The idea is to extract, from the obtained
numerical solution $\widehat{A}^{\;\mu}(\tau)$
at an equilibrium time $\tau_\text{eq}$,
the constant master-field matrices $\widehat{a}^{\;\mu}$ by use
of \eqref{eq:IIB-matrix-model-master-field},
for the given values of $\widehat{p}_{k}$.
The obtained matrices $\widehat{a}^{\;\mu}$ are expected to solve
the algebraic equation \eqref{eq:IIB-matrix-model-algebraic-equation},
at least within the numerical accuracy.
Assuming that the trace term \eqref{eq:varPfoverPf-equals-trace}  
can be evaluated accurately (with advanced numerical methods
and powerful computers), it would seem that   
master-field matrices $\widehat{a}^{\;\mu}$
with sizes $N=64$ and $N=128$ could be within reach.

\newpage   %%tmp
\begin{appendix}

\section{Pfaffian for $D=10$ and $N \geq 2$}
\label{app:Pfaffian-D10-Ngeq2}

The fermion integration in the partition function
of the $D=10$ IIB matrix model \eqref{eq:IIB-matrix-model-Z}
with fermionic action \eqref{eq:Sferm}
gives the Pfaffian \eqref{eq:Pfaffian-calP}
as a function of the bosonic matrices $A_{\mu}$.
Explicitly, the Pfaffian evaluated for the master-field
matrices $\widehat{a}_{\mu}$
reads~\cite{KrauthNicolaiStaudacher1998,NishimuraVernizzi2000-JHEP}
\bsubeqs\label{eq:calP-calM}
\beqa
\label{eq:calP}
&&
\mathcal{P}\left(\widehat{a}\,\right) \equiv
\text{Pf}\left[\mathcal{M}\left(\widehat{a}\,\right)\right]\,,
\\[2mm]
\label{eq:calM}
&&
\mathcal{M}_{a\alpha\,,\,b\beta}
 =
-i\, f_{abc}\, \big(\mathcal{C}\,
\Sigma_{\mu}\big)_{\alpha\beta} \;
\widehat{a}_{\mu}^{\;c}\,,
\eeqa
\esubeqs
with Lie-algebra indices $a,b,c$ running over $\{1,\, \ldots \,,\, (N^2-1)\}$,
spinorial indices $\alpha,\beta$ running over $\{1,\, \ldots \,,\, 16\}$,
and the directional index $\mu$ being summed
over $\{  1,\,  \ldots \,,\,10 \}$,
where the pair of indices $a\alpha$ gives a combined index $A$
and $b\beta$ gives a combined index $B$.
For completeness, we give the definition of the Pfaffian  
in terms of the completely antisymmetric Levi--Civita symbol $\epsilon$
normalized to $1$. The Pfaffian of
a $(2 n)\times (2 n)$ skew-symmetric matrix $A=(a_{ij})$
is then given by
\beq
\label{eq:def-Pfaffian}
\text{Pf}[A]\equiv
\frac{1}{2^{n}\,n!}\;
\epsilon_{i_{1}j_{1} i_{2}j_{2} \cdots i_{n}j_{n} }\;
a_{i_{1}j_{1}}\,a_{i_{2}j_{2}} \cdots a_{i_{n}j_{n}}\,,
\eeq
with implicit summations over repeated indices.

The real numbers $f_{abc}$ in \eqref{eq:calM} are the structure constants
from the $SU(N)$ traceless Hermitian generators $t_{a}$: 
\bsubeqs\label{eq:fabc-TraceNormalization}
\beqa\label{eq:fabc}
&&
f_{abc}
=
-2\, i\, \text{Tr}\, \big( t_{a}\, \big[ t_{b} ,\, t_{c} \big] \big)\,,
\\[2mm]
\label{eq:TraceNormalization}
&&
\text{Tr}\, \big( t_{a} \cdot t_{b} \big)
=
\frac{1}{2}\;\delta_{a b}\,,
\eeqa
\esubeqs
where the last equation sets the normalization of the generators.
The coefficients $\widehat{a}_{\mu}^{\;c}$ in \eqref{eq:calM}
have resulted from the expansion of the bosonic master-field
matrix $\widehat{a}_{\mu}$ with respect to these generators, 
\beqa
\label{eq: master-field-matrix-expansion}
&&
\left(\widehat{a}_{\mu}\right)_{kl}
=\widehat{a}_{\mu}^{\;c}\,\left(t_{c}\right)_{kl} \,.
\eeqa
The charge conjugation matrix $\mathcal{C}$ in \eqref{eq:calM}
becomes the unit matrix for an appropriate choice of
$\Sigma_{\mu}$ matrices~\cite{NishimuraVernizzi2000-JHEP}: 
\beq\label{eq:Sigma-mu}
\begin{array}{lll}
\hspace*{-0mm}
&\Sigma_{1\phantom{0}}
=\; i\,\sigma_{2}\otimes\sigma_{2}\otimes\sigma_{2}\otimes\sigma_{2}\,,
\quad
&\Sigma_{2\phantom{0}}
=\;  i\,\sigma_{2}\otimes\sigma_{2}\otimes\id\otimes\sigma_{1}\,,
\\[2mm]
&\Sigma_{3\phantom{0}}
=\;  i\,\sigma_{2}\otimes\sigma_{2}\otimes\id\otimes\sigma_{3}\,,
\quad
&\Sigma_{4\phantom{0}}
=\; i\,\sigma_{2}\otimes\sigma_{1}\otimes\sigma_{2}\otimes\id\,,
\\[2mm]
&\Sigma_{5\phantom{0}}
=\;  i\,\sigma_{2}\otimes\sigma_{3}\otimes\sigma_{2}\otimes\id\,,
\quad\;\;
&\Sigma_{6\phantom{0}}
=\;  i\,\sigma_{2}\otimes\id\otimes\sigma_{1}\otimes\sigma_{2}\,,
\\[2mm]
\hspace*{-0mm}
&\Sigma_{7\phantom{0}}
 =\; i\,\sigma_{2}\otimes\id\otimes\sigma_{3}\otimes\sigma_{2}\,,
\quad
&\Sigma_{8\phantom{0}}
=\;  i\,\sigma_{1}\otimes\id\otimes\id\otimes\id\,,
\\[2mm]
&\Sigma_{9\phantom{0}}
=\;  i\,\sigma_{3}\otimes\id\otimes\id\otimes\id\,,
\quad
&\Sigma_{10}
=\; \id\otimes\id\otimes\id\otimes\id \;=\; \id_{16}  \,,
\\
\end{array}
\eeq
which are manifestly symmetric
(so that the corresponding matrix $\mathcal{C}$ is trivial). 
Incidentally, we prefer to call these matrices ``Sigma''
by analogy to the 4-dimensional case
with the four $4\times 4$ Dirac matrices $\gamma_{\mu}$
and the three $2\times 2$ Pauli matrices $\sigma_{a}$,
to which $\sigma_{4} \equiv \id$ is added.

The matrix dimension of $\mathcal{M}$ equals 48 for $N=2$,
128 for $N=3$, and 240 for $N=4$.
With arbitrary numerical coefficients $\widehat{a}_{\mu}^{\;c}$,
it can be readily verified that
$\det[\mathcal{M}] =(\text{Pf}[\mathcal{M}])^2$
is positive and real for $N=2$ and $3$, but complex
for $N=4$. This implies that the above Pfaffian is
real for  $N=2$ and $3$ and complex for $N=4$.
The above Pfaffian is, in general, complex
also for $N>4$~\cite{NishimuraVernizzi2000-JHEP}.

The homogeneous polynomial $\mathcal{P}\left(\widehat{a}\,\right)$
from \eqref{eq:calP} has an order equal to half the
dimension of $\mathcal{M}\left(\widehat{a}\,\right)$. Hence,
the polynomial $\mathcal{P}\left(\widehat{a}\,\right) $ has order 24, 64, and 120
for $N=2$, $3$, and $4$, respectively.
The corresponding algebraic equations \eqref{eq:full-algebraic-equation}
are indeed huge.

%%%%%%%\newpage  %%tmp

\section{Pseudorandom numbers for $(D,\,N)=(10,\,4)$ }
\label{app:Pseudorandom numbers}

In this appendix, we give the
particular realization (the ``$\kappa$-realization'') of
the pseudorandom numbers used for the
approximate numerical solution of Sec.~\ref{subsec:Case-D10-N4} .

Specifically, we take the following 4 real pseudorandom numbers  
for the master momenta:
\beqa
\label{eq:phat-kappa-realization} 
\hspace*{-11mm}&&
\widehat{p}_\text{$\kappa$-realization}
=
\left\{{-\frac{111}{250},\,\frac{19}{200},\,
       -\frac{63}{200},\,\frac{189}{1000}}\right\}\,, %%\scriptstyle
\eeqa
and the following 150 real pseudorandom numbers entering the
Hermitian master-noise matrices:

\bsubeqs\label{eq:etahat-kappa-realization}
\beqa
\hspace*{-14mm}&&
\widehat{\eta}^{\,1}_\text{$\kappa$-realization}=
\nonumber\\[1mm]
\hspace*{-14mm}&&
\renewcommand{\arraystretch}{1.25}  %% enlarge line spacing
\left(\!\!
\begin{array}{cccc}
 \frac{593}{2000 \sqrt{2}}-\frac{151}{500} & \frac{1}{2000}-\frac{9 i}{40} &
 \frac{353}{2000}+\frac{6 i}{25} & -\frac{987}{2000}-\frac{51 i}{400} \\
 \frac{1}{2000}+\frac{9 i}{40} & \frac{151}{500}+\frac{593}{2000 \sqrt{2}} &
 \frac{1}{8}-\frac{63 i}{2000} & -\frac{131}{400}-\frac{367 i}{2000} \\
 \frac{353}{2000}-\frac{6 i}{25} & \frac{1}{8}+\frac{63 i}{2000} &
 -\frac{369}{1000}-\frac{593}{2000 \sqrt{2}} & -\frac{169}{400}+\frac{171 i}{500} \\
 -\frac{987}{2000}+\frac{51 i}{400} & -\frac{131}{400}+\frac{367 i}{2000} &
 -\frac{169}{400}-\frac{171 i}{500} & \frac{369}{1000}-\frac{593}{2000 \sqrt{2}} \\
\end{array}
\!\!\right)\!,
\eeqa
%%\\[1mm]
\beqa
\hspace*{-14mm}&&
\widehat{\eta}^{\,2}_\text{$\kappa$-realization}=
\nonumber\\[1mm]\hspace*{-14mm}&&
\renewcommand{\arraystretch}{1.25}  %% enlarge line spacing
\left(\!\!
\begin{array}{cccc}
 \frac{69}{2000}-\frac{17}{50 \sqrt{2}} & -\frac{153}{500}-\frac{47 i}{500} &
 \frac{897}{2000}-\frac{61 i}{1000} & \frac{269}{2000}-\frac{237 i}{2000} \\
 -\frac{153}{500}+\frac{47 i}{500} & -\frac{69}{2000}-\frac{17}{50 \sqrt{2}} &
 \frac{1}{250}-\frac{13 i}{200} & -\frac{103}{400}+\frac{367 i}{1000} \\
 \frac{897}{2000}+\frac{61 i}{1000} & \frac{1}{250}+\frac{13 i}{200} & \frac{17}{50
 \sqrt{2}}-\frac{123}{250} & \frac{1}{20}-\frac{71 i}{2000} \\
 \frac{269}{2000}+\frac{237 i}{2000} & -\frac{103}{400}-\frac{367 i}{1000} &
 \frac{1}{20}+\frac{71 i}{2000} & \frac{123}{250}+\frac{17}{50 \sqrt{2}} \\
\end{array}
\!\!\right)\!,
\eeqa

\beqa
\hspace*{-14mm}&&
\widehat{\eta}^{\,3}_\text{$\kappa$-realization}=
\nonumber\\[1mm]\hspace*{-14mm}&&
\renewcommand{\arraystretch}{1.25}  %% enlarge line spacing
\left(\!\!
\begin{array}{cccc}
 \frac{313}{1000}+\frac{7}{250 \sqrt{2}} & \frac{7}{200}+\frac{431 i}{2000} &
 \frac{17}{40}-\frac{59 i}{250} & -\frac{437}{2000}+\frac{69 i}{1000} \\
 \frac{7}{200}-\frac{431 i}{2000} & \frac{7}{250 \sqrt{2}}-\frac{313}{1000} &
 -\frac{991}{2000}-\frac{49 i}{250} & \frac{83}{1000}-\frac{279 i}{2000} \\
 \frac{17}{40}+\frac{59 i}{250} & -\frac{991}{2000}+\frac{49 i}{250} &
 \frac{871}{2000}-\frac{7}{250 \sqrt{2}} & \frac{49}{250}+\frac{121 i}{500} \\
 -\frac{437}{2000}-\frac{69 i}{1000} & \frac{83}{1000}+\frac{279 i}{2000} &
 \frac{49}{250}-\frac{121 i}{500} & -\frac{871}{2000}-\frac{7}{250 \sqrt{2}} \\
\end{array}
\!\!\right)\!,
\eeqa
%%\\[1mm]
\beqa
\hspace*{-14mm}&&
\widehat{\eta}^{\,4}_\text{$\kappa$-realization}=
\nonumber\\[1mm]\hspace*{-14mm}&&
\renewcommand{\arraystretch}{1.25}  %% enlarge line spacing
\left(\!\!
\begin{array}{cccc}
 \frac{293}{1000}+\frac{429}{2000 \sqrt{2}} & -\frac{11}{250}-\frac{469 i}{2000} &
 -\frac{439}{1000}-\frac{483 i}{2000} & -\frac{11}{50}+\frac{41 i}{400} \\
 -\frac{11}{250}+\frac{469 i}{2000} & \frac{429}{2000 \sqrt{2}}-\frac{293}{1000} &
 \frac{409}{1000}-\frac{343 i}{2000} & \frac{3}{16}-\frac{407 i}{1000} \\
 -\frac{439}{1000}+\frac{483 i}{2000} & \frac{409}{1000}+\frac{343 i}{2000} &
 -\frac{6}{125}-\frac{429}{2000 \sqrt{2}} & -\frac{141}{500}-\frac{31 i}{2000} \\
 -\frac{11}{50}-\frac{41 i}{400} & \frac{3}{16}+\frac{407 i}{1000} & -\frac{141}{500}+\frac{31
 i}{2000} & \frac{6}{125}-\frac{429}{2000 \sqrt{2}} \\
\end{array}
\!\!\right)\!,
\eeqa

\beqa
\hspace*{-14mm}&&
\widehat{\eta}^{\,5}_\text{$\kappa$-realization}=
\nonumber\\[1mm]\hspace*{-14mm}&&
\renewcommand{\arraystretch}{1.25}  %% enlarge line spacing
\left(\!\!
\begin{array}{cccc}
 \frac{183}{500 \sqrt{2}}-\frac{48}{125} & \frac{37}{125}+\frac{249 i}{1000} &
 -\frac{511}{2000}-\frac{43 i}{500} & -\frac{129}{2000}+\frac{19 i}{200} \\
 \frac{37}{125}-\frac{249 i}{1000} & \frac{48}{125}+\frac{183}{500 \sqrt{2}} &
 \frac{49}{400}-\frac{463 i}{1000} & \frac{23}{200}+\frac{433 i}{2000} \\
 -\frac{511}{2000}+\frac{43 i}{500} & \frac{49}{400}+\frac{463 i}{1000} &
 -\frac{46}{125}-\frac{183}{500 \sqrt{2}} & \frac{821}{2000}+\frac{199 i}{2000} \\
 -\frac{129}{2000}-\frac{19 i}{200} & \frac{23}{200}-\frac{433 i}{2000} &
 \frac{821}{2000}-\frac{199 i}{2000} & \frac{46}{125}-\frac{183}{500 \sqrt{2}} \\
\end{array}
\!\!\right)\!,
\eeqa
%%\\[1mm]
\beqa
\hspace*{-14mm}&&
\widehat{\eta}^{\,6}_\text{$\kappa$-realization}=
\nonumber\\[1mm]\hspace*{-14mm}&&
\renewcommand{\arraystretch}{1.25}  %% enlarge line spacing
\left(\!\!
\begin{array}{cccc}
 \frac{181}{1000}-\frac{181}{2000 \sqrt{2}} & -\frac{143}{500}+\frac{i}{5} &
 \frac{921}{2000}-\frac{313 i}{1000} & -\frac{603}{2000}-\frac{449 i}{2000} \\
 -\frac{143}{500}-\frac{i}{5} & -\frac{181}{1000}-\frac{181}{2000 \sqrt{2}} &
 \frac{609}{2000}-\frac{39 i}{2000} & \frac{443}{2000}+\frac{383 i}{1000} \\
 \frac{921}{2000}+\frac{313 i}{1000} & \frac{609}{2000}+\frac{39 i}{2000} &
 \frac{941}{2000}+\frac{181}{2000 \sqrt{2}} & -\frac{17}{1000}+\frac{27 i}{2000} \\
 -\frac{603}{2000}+\frac{449 i}{2000} & \frac{443}{2000}-\frac{383 i}{1000} &
 -\frac{17}{1000}-\frac{27 i}{2000} & \frac{181}{2000 \sqrt{2}}-\frac{941}{2000} \\
\end{array}
\!\!\right)\!,
\eeqa

\beqa
\hspace*{-14mm}&&
\widehat{\eta}^{\,7}_\text{$\kappa$-realization}=
\nonumber\\[1mm]\hspace*{-14mm}&&
\renewcommand{\arraystretch}{1.25}  %% enlarge line spacing
\left(\!\!
\begin{array}{cccc}
 \frac{57}{2000}+\frac{83}{200 \sqrt{2}} & -\frac{129}{500}-\frac{147 i}{2000} &
 \frac{387}{2000}-\frac{611 i}{2000} & \frac{313}{2000}+\frac{191 i}{400} \\
 -\frac{129}{500}+\frac{147 i}{2000} & \frac{83}{200 \sqrt{2}}-\frac{57}{2000} &
 -\frac{61}{500}-\frac{11 i}{50} & -\frac{969}{2000}+\frac{927 i}{2000} \\
 \frac{387}{2000}+\frac{611 i}{2000} & -\frac{61}{500}+\frac{11 i}{50} &
 -\frac{2}{125}-\frac{83}{200 \sqrt{2}} & \frac{489}{1000}+\frac{361 i}{1000} \\
 \frac{313}{2000}-\frac{191 i}{400} & -\frac{969}{2000}-\frac{927 i}{2000} &
 \frac{489}{1000}-\frac{361 i}{1000} & \frac{2}{125}-\frac{83}{200 \sqrt{2}} \\
\end{array}
\!\!\right)\!,
\eeqa
%%\\[1mm]
\beqa
\hspace*{-14mm}&&
\widehat{\eta}^{\,8}_\text{$\kappa$-realization}=
\nonumber\\[1mm]\hspace*{-14mm}&&
\renewcommand{\arraystretch}{1.25}  %% enlarge line spacing
\left(\!\!
\begin{array}{cccc}
 \frac{321}{2000}-\frac{193}{2000 \sqrt{2}} & -\frac{273}{2000}-\frac{441 i}{2000} &
 \frac{427}{1000}-\frac{317 i}{2000} & -\frac{407}{2000}-\frac{57 i}{400} \\
 -\frac{273}{2000}+\frac{441 i}{2000} & -\frac{321}{2000}-\frac{193}{2000 \sqrt{2}} &
 \frac{23}{2000}+\frac{93 i}{400} & \frac{163}{500}+\frac{18 i}{125} \\
 \frac{427}{1000}+\frac{317 i}{2000} & \frac{23}{2000}-\frac{93 i}{400} &
 \frac{307}{1000}+\frac{193}{2000 \sqrt{2}} & \frac{333}{2000}-\frac{9 i}{25} \\
 -\frac{407}{2000}+\frac{57 i}{400} & \frac{163}{500}-\frac{18 i}{125} &
 \frac{333}{2000}+\frac{9 i}{25} & \frac{193}{2000 \sqrt{2}}-\frac{307}{1000} \\
\end{array}
\!\!\right)\!,
\eeqa

\beqa
\hspace*{-14mm}&&
\widehat{\eta}^{\,9}_\text{$\kappa$-realization}=
\nonumber\\[1mm]\hspace*{-14mm}&&
\renewcommand{\arraystretch}{1.25}  %% enlarge line spacing
\left(\!\!
\begin{array}{cccc}
 -\frac{39}{400}-\frac{7}{200 \sqrt{2}} & \frac{1}{500}-\frac{i}{125} &
 -\frac{981}{2000}+\frac{56 i}{125} & -\frac{137}{1000}+\frac{547 i}{2000} \\
 \frac{1}{500}+\frac{i}{125} & \frac{39}{400}-\frac{7}{200 \sqrt{2}} &
 -\frac{201}{1000}+\frac{101 i}{1000} & -\frac{63}{500}+\frac{31 i}{125} \\
 -\frac{981}{2000}-\frac{56 i}{125} & -\frac{201}{1000}-\frac{101 i}{1000} &
 \frac{197}{400}+\frac{7}{200 \sqrt{2}} & \frac{597}{2000}-\frac{57 i}{125} \\
 -\frac{137}{1000}-\frac{547 i}{2000} & -\frac{63}{500}-\frac{31 i}{125} &
 \frac{597}{2000}+\frac{57 i}{125} & \frac{7}{200 \sqrt{2}}-\frac{197}{400} \\
\end{array}
\!\!\right)\!,
\eeqa
%%\\[1mm]
\beqa
\hspace*{-14mm}&&
\widehat{\eta}^{\,10}_\text{$\kappa$-realization}=
\nonumber\\[1mm]\hspace*{-14mm}&&
\renewcommand{\arraystretch}{1.25}  %% enlarge line spacing
\left(\!\!
\begin{array}{cccc}
 \frac{577}{2000 \sqrt{2}}-\frac{319}{2000} & \frac{969}{2000}+\frac{2 i}{5} &
 -\frac{243}{2000}+\frac{29 i}{500} & -\frac{19}{125}-\frac{151 i}{1000} \\
 \frac{969}{2000}-\frac{2 i}{5} & \frac{319}{2000}+\frac{577}{2000 \sqrt{2}} &
 \frac{17}{50}+\frac{119 i}{2000} & \frac{463}{2000}-\frac{33 i}{400} \\
 -\frac{243}{2000}-\frac{29 i}{500} & \frac{17}{50}-\frac{119 i}{2000} &
 \frac{119}{1000}-\frac{577}{2000 \sqrt{2}} & \frac{257}{1000}+\frac{219 i}{500} \\
 -\frac{19}{125}+\frac{151 i}{1000} & \frac{463}{2000}+\frac{33 i}{400} &
 \frac{257}{1000}-\frac{219 i}{500} & -\frac{119}{1000}-\frac{577}{2000 \sqrt{2}} \\
\end{array}
\!\!\right)\!.
\vspace*{0mm}
\eeqa
\esubeqs
%\noindent
\vspace*{0.25\baselineskip}

Incidentally, the above matrices \eqref{eq:etahat-kappa-realization}
also make clear what $4 \times 4$ matrix realizations have been chosen    
for the
$SU(4)$ generators $t_{a}$ from \eqref{eq:fabc-TraceNormalization}.

\end{appendix}

%%%%%%%%%%%%%%%%%%%%%%\newpage


\begin{thebibliography}{99}


\bibitem{Witten1979}
E.~Witten,
``The $1/N$ expansion in atomic and particle physics,''
in G. 't Hooft et. al (eds.),
\emph{Recent Developments in Gauge Theories},
Cargese 1979 (Plenum Press, New York, 1980).



\bibitem{GreensiteHalpern1983}
J.~Greensite and M.B.~Halpern,
\hspace*{0mm}``Quenched master fields,''
Nucl. Phys. B \textbf{211}, 343 (1983).
%%doi:10.1016/0550-3213(83)90413-3

\bibitem{IKKT-1997}
N.~Ishibashi, H.~Kawai, Y.~Kitazawa, and A.~Tsuchiya,
``A large-$N$ reduced model as superstring,''
Nucl.\ Phys.\ B \textbf{498}, 467 (1997),
arXiv:hep-th/9612115.
  %%CITATION = doi:10.1016/S0550-3213(97)00290-3;%%



\bibitem{Aoki-etal-review-1999}
H.~Aoki, S.~Iso, H.~Kawai, Y.~Kitazawa, A.~Tsuchiya, and T.~Tada,
``IIB matrix model,''
Prog.\ Theor.\ Phys.\ Suppl.\  \textbf{134}, 47 (1999),
arXiv:hep-th/9908038.
%%CITATION = doi:10.1143/PTPS.134.47;%%


\bibitem{Klinkhamer2020-master}
F.R.~Klinkhamer,
``IIB matrix model: Emergent spacetime from the master field,''
Prog. Theor. Exp. Phys. \textbf{2021}, 013B04 (2021),
arXiv:2007.08485. %% [hep-th]].




\bibitem{Klinkhamer2020-reg-bb-IIB-m-m}
F.R.~Klinkhamer,
``IIB matrix model and regularized big bang,''
Prog. Theor. Exp. Phys. \textbf{2021}, 063B05 (2021),
arXiv:2009.06525. %% [hep-th]].


\bibitem{Klinkhamer2021-APPB-review}
F.R.~Klinkhamer,
``M-theory and the birth of the Universe,''
Acta Phys. Pol. B \textbf{52}, 1007  (2021),
arXiv:2102.11202. %% [hep-th]].



\bibitem{Klinkhamer2021-first-look}
F.R.~Klinkhamer,
``A first look at the bosonic master-field equation of the IIB matrix model,''
Int. J. Mod. Phys. D \textbf{30}, 2150105 (2021)
arXiv:2105.05831.  %% [hep-th]].


\bibitem{Klinkhamer2021-sols-D3-N3}
F.R.~Klinkhamer,
``Solutions of the bosonic master-field equation
from a supersymmetric matrix model,''
Acta Phys. Pol. B \textbf{52}, 1339  (2021),
arXiv:2106.07632.  %% [hep-th]].


\bibitem{KimNishimuraTsuchiya2012}
S.W.~Kim, J.~Nishimura, and A.~Tsuchiya,
\hspace*{0mm}``Expanding (3+1)-dimensional universe from a Lorentzian matrix model
\hspace*{0mm}  for superstring theory in (9+1)-dimensions,''
  Phys.\ Rev.\ Lett.\  {\bf 108}, 011601 (2012),
arXiv:1108.1540. %% [hep-th]].
%%CITATION = doi:10.1103/PhysRevLett.108.011601;%%



\bibitem{NishimuraTsuchiya2019}
J.~Nishimura and A.~Tsuchiya,
\hspace*{0mm}``Complex Langevin analysis of the space-time structure
\hspace*{0mm}  in the Lorentzian type IIB matrix model,''
JHEP {\bf 1906}, 077 (2019),
arXiv:1904.05919. %% [hep-th]].
  %%CITATION = doi:10.1007/JHEP06(2019)077;%%



\bibitem{Anagnostopoulos-etal-2020}
K.N.~Anagnostopoulos, T.~Azuma, Y.~Ito, J.~Nishimura, T.~Okubo,
and S.~K. Papadoudis,
``Complex Langevin analysis of the spontaneous breaking
of 10D rotational symmetry in the Euclidean IKKT matrix model,''
JHEP \textbf{2006}, 069 (2020),
%%doi:10.1007/JHEP06(2020)069
arXiv:2002.07410.  %%[hep-th]].

\bibitem{KrauthNicolaiStaudacher1998}
W.~Krauth, H.~Nicolai, and M.~Staudacher,
``Monte Carlo approach to $M$-theory,''
Phys. Lett. B \textbf{431}, 31 (1998),
%%doi:10.1016/S0370-2693(98)00557-7
arXiv:hep-th/9803117.


\bibitem{NishimuraVernizzi2000-JHEP}
J.~Nishimura and G.~Vernizzi,
``Spontaneous breakdown of Lorentz invariance in IIB matrix model,''
JHEP \textbf{0004}, 015 (2000),
%%doi:10.1088/1126-6708/2000/04/015
arXiv:hep-th/0003223.  %%[hep-th]].

\bibitem{Nishimura-private}
J.~Nishimura, private communication (2021).


\bibitem{Wolfram1991}
S.~Wolfram,
\emph{Mathematica: A System for Doing Mathematics by Computer,
Second Edition}
(Addison--Wesley, Redwood City CA, USA, 1991).


\bibitem{NelderMead1965}
J.A. Nelder and R. Mead,
``A simplex method for function minimization,''
The Computer Journal, {\bf 7}, 308 (1965);
Errata \textit{ibid}, {\bf 8}, 27 (1965).
%https://doi.org/10.1093/comjnl/7.4.308
%https://doi.org/10.1093/comjnl/8.1.27


\bibitem{Press-etal-1992}
W.H.~Press, S.A.~Teukolsky, W.T.~Vetterling, and B.P.~Flannery,
\emph{Numerical Recipes in FORTRAN: The Art of Scientific Computing}
(Cambridge University Press, Cambridge, UK, 1986).


\bibitem{Klinkhamer2021-Corfu-v4}
F.R.~Klinkhamer,
``IIB matrix model, bosonic master field, and emergent spacetime,''
talk at: \textit{Workshop on Quantum Geometry, Field Theory, and Gravity},
Corfu Summer Institute, September 20--27, 2021
[slides available from \verb"https://www.itp.kit.edu/_media/research/corfu2021-klinkhamer-v4."\newline
\verb"pdf"]. 

\end{thebibliography}
\end{document}